\documentclass[12pt,a4paper]{article}
\usepackage{amsmath,amssymb}
\usepackage{exscale}
\usepackage{relsize}%\mathlarger\int

\def\endproof{\par\nobreak\hbox to \hsize{\hfil\vrule width 5pt height
5pt}\goodbreak\vskip 3pt}

\begin{document}

\title{Reconstructing Adiabats}
\author{J. B. Cooper\\ Johannes Kepler Universit\"at Linz}
\date{}
\maketitle

%\tableofcontents
%\newpage
\begin{abstract}
We give an explicit method for calculating the equations of state of a thermodynamical substance from the following information: in the calibrated case,
the system of isotherms and
a {\it single} adiabat, in the uncalibrated one, the same but with {\it two}
adiabats.  For simple examples the entropy function can be computed explicitly from this material but it involves steps (solution of 
non-linear equations  and an integration) which will usually
require numerical methods for their implementation.

\end{abstract}

\section{Introduction}

The equations of state of a classical Gibbsian thermodynamical system
are of the form
$$T = f(p,V), \quad S = g(p,V)$$
where $p$, $V$, $T$ and $S$ are the pressure, volume, (absolute) temperature and (absolute) entropy respectively.
The isotherms and adiabats are then the level curves of $f$ and $g$.
In [Co] we considered the question of how much information is required to
obtain these equations for a given substance.  We showed
that if one knows $T$ as the  function $f$ of $p$ and $V$
and {\it one} adiabat, i.e.,
a single level curve of $g$, then one can recover $g$.  If, on the other hand,
one knows only the level curves of $T$ (the isotherms), then {\it two}
adiabats are required.

The reason why this is possible is  that $f$ and $g$ are locked into a 
mutual dependence by the so-called Maxwell relations, one
version of which states that the Jacobi determinant
$$\frac{\partial (T,S)}{\partial (p,V)} $$
is identically $1$ which means that 
the mapping from $(p,V)$-space into $(T,S)$-space
is area-preserving.
 
In this short article we give a \lq \lq theory light'' version 
in the form of a step by step method for computing the adiabats 
from the above information.

We begin with a remark on notation.  For our purposes it would be rather 
confusing to employ
the standard thermodynamical variables and  
we shall use the  mathematically  neutral letters $x$ and $y$ (which 
correspond to $p$
and $V$ in Gibbsian thermodynamics) 
etc. for
our variables.

\section{Recovering adiabats---the recipe}
The general strategy of our method is a standard one in physics---we
introduce new coordinates which preserve the essential elements of the
situation and reduce the problem to a more tractable one, in this case 
where the
isotherms are parallel lines.

We suppose that we are given the equation of state $T = f(p,V)$ and one
adiabat. 
We use, as mentioned above, 
$x$ and $y$ as the independent variables.  We construct in six steps
two coordinate transformations with $(X,Y)$ and 
$(\widetilde X,\widetilde Y)$ 
for the resulting variables.

\noindent Step 1.  We introduce new variables

$$X = f(x,y), \quad  Y = x.$$
 
\noindent Step 2.
We invert the above transformation to get $x$ and $y$ as functions of $X$
and $Y$.

\noindent Step 3.  We define a function $\Psi$ as follows:
 $$\Psi(X,Y) =- \int \frac 1{f_y}\, dY = -\int \frac{dY}{f_y(x(X,Y),y(X,Y))}.$$

\noindent Step 4.
We then define further new variables
 $$\widetilde X = X, \quad \widetilde Y = \Psi(X,Y).$$
Again, we invert the transformation to get  $(x,y)$ as functions of
$\widetilde X$ and
$\widetilde Y$. 

\noindent Step 5.
We consider the one given adiabat which we assume is defined implicitly
by the equation $g(x,y)=0$.  Then this curve is transformed into the
equation  $h(\widetilde X,\widetilde Y) = 0$ in the 
$(\widetilde X,\widetilde Y)$-plane
where 
$$h(\widetilde X,\widetilde Y)= g(x(\widetilde X,\widetilde Y),y(\widetilde X,
\widetilde Y)).$$
We write the equation of this curve as the graph of a function
 in the form  $\widetilde Y=f(\widetilde X)$ by solving 
$h(\widetilde X,\widetilde Y)=0$ for $\widetilde Y$.
Then the  entropy function in the transformed plane is 
$\widetilde Y -f(\widetilde X)$ and so
the adiabats in $(\widetilde X,\widetilde Y)$-space are the translates
of the given one.  

\noindent Step 6.  The adiabats in $(x,y)$-space are then the transforms
of this family, i.e., 
the entropy function is $\widetilde Y(x,y) -f(\widetilde X(x,y))$ and 
the adiabats are its level
curves.

\vskip5pt \noindent In order to illustrate this process, we begin with a  
simple example---the ideal gas where each of the above steps 
can be carried out explicitly:  
of course, in this case we know what the result should be but it 
is interesting to see how the method pulls the familiar logarithmic term in 
the formula
for the entropy out of the hat.

Our starting point is the equation $X = x y$ which corresponds to Charles'
law $T=pV$  for the ideal gas (we omit constants).
We assume that we have empirical data on {\it one} adiabat which
we model with a  generalised parabola $y=x^\alpha$.
We now show that this information suffices to reveal the entropy function
and so the family of {\it all} adiabats.

\noindent Step 1.  The first transformation is
 $X = x y$,  $Y = x$.

\noindent Step 2.
We solve for $x$ and $y$
to get $x = Y$ and $y = \dfrac X Y$.  

\noindent Step 3.
For  $\Psi$ we have
$$\Psi(X,Y) = -\int \frac 1 x \,dY = -\int \dfrac 1 Y \,dY = -\ln Y.$$

\noindent Step 4.
Putting this together, we have $\widetilde X = x y, \quad \widetilde Y = - \ln x$.

\noindent Step 5.
We can invert these equations to get 
$$x = e^{-\widetilde Y}, \quad y = \widetilde X e^{\widetilde Y}. $$

Then the equation $y = x^\alpha$ is transformed to the form 
$\widetilde Y + \gamma \ln \widetilde X + \gamma \widetilde Y = 0$ where 
$\gamma = - \dfrac 1 \alpha$ and so 
the adiabat
is the graph of the function 
$\widetilde Y=-\frac \gamma{\gamma-1}\ln \widetilde X$. 
Translating this function back into one in $x$ and $y$ 
gives the familiar entropy function
$$S=\dfrac 1{\gamma-1}\ln(x y^\gamma)$$ 
with the corresponding level curves.

\vskip5pt
This example was chosen for its familiarity and simplicity which
allows us to  carry out the necessary computations explicitly and by hand.

In practical situations we can expect to meet difficulties

\noindent a) in inverting the transformations from the $x$ and $y$ coordinates 
to $X$ and $Y$ and then to $\widetilde X$ and $\widetilde Y$---this 
involves solving  
pairs of non-linear equations 
for $x$ and $y$;

 \noindent b) in computing the integral used to define $\Psi$;

\noindent c) in expressing the transform of the given adiabat in the
form of the graph of a function---this, again, involves solving a
non-linear equation.

\noindent For real gases,  some of 
these processes may have to be computed with the aid
of numerical methods.

\vskip5pt \noindent 
In order to show the practical difficulties which can arise in real gases, we 
consider the case where $X = \phi(x y)$ for some calibrating function
$\phi$. This  means that Boyle's law but not Charles' holds.

We find this example particularly interesting since it treats a phenomenon
which occurs frequently in practice but which is scarely dealt with in the
textbooks, namely one where we have to recalibrate temperature. This means
that the 
first equation of state has the form $T = \phi(p V)$ for some non-trivial
calibration function $\phi$.

An example of this phenomenon is one due to Feynman [Fe]---we
call it the  Feynman gas.
Here $t= p V$ and $s=p V^{\gamma(t)}$ for a function $\gamma$
of one variable (we are using lower case letters $t$ and $s$ to indicate that
these are empirical temperature and entropy respectively). 
This example was introduced by Feynman (see [Fe])
to cope with the fact that, in a real gas, the adiabatic index depends on 
temperature. 
This is the first example which we have met in the literature
where we genuinely have
to recalibrate temperature (i.e., Boyle's law holds only in the weak form
that $pV$ is constant for constant temperature). This fact is not considered
by Feynman but follows from the analysis in [Co]
where it was shown that absolute temperature is then given by the equation
$T =\phi(t) =\phi(p V)$
where $\phi$ is a primitive
of $\dfrac 1{\gamma-1}$ 
(recall that $\gamma$ is now a function of one variable).
  
For a discussion of the relevance of 
such recalibrations, see Chang [Ch].  The recalibrations introduced here
provide
an at least qualitative explanation of the diagram on p. 78 of this reference,
displaying comparative data of Le Duc on spirit thermometers.

\vskip5pt\noindent
We return to our example and compute:

$$\Psi(X,Y) = - \frac1{\phi'(\phi^{-1}(X))} \ln Y$$
and hence $\widetilde X  = \phi(x y)$,
$\widetilde Y = -\ln x \phi'(x y)$.

We then have $$x =\exp\left(-\frac{\widetilde Y}{\phi'(\phi^{-1}(\widetilde X)}\right),\quad  y=\frac{\phi^{-1}(\widetilde X)}{\exp\left(-\frac{\widetilde Y}{\phi'(\phi^{-1}(\widetilde X)}\right)}.$$
Once again, we assume that we have one adiabat given in  implicit 
form $g(x,y) = 0$.

Hence the given adiabat transfers to the form
$$g\left(\exp\left(-\frac{\widetilde Y}{\phi'(\phi^{-1}(\widetilde X)}\right),\quad 
\phi^{-1}(\widetilde X) \exp\left(\frac{\widetilde Y}{\phi'(\phi^{-1}(\widetilde X)}\right)\right)=0. $$  In this case, we will generally not be able to solve this
explicitly for $\widetilde Y$ and so will have to employ numerical methods to
complete the analysis.

\section{Why does this work?}

The background for the above method is the well-known fact (perhaps not
as well-known as it should be) that each of the four Maxwell relations in 
thermodynamics is (individually) equivalent to
the Jacobian identity 
$\dfrac {\partial (S,T)}{\partial (p,V)} = 1$ (see, for example, [Ri]).  
This
means that the the corresponding map from the  $(p,V)$-plane
into the $(S,T)$-plane is area-preserving.

Roughly speaking this says that much of the thermodynamical structure remains
intact if we subject the variables to an area-preserving transformation of 
the coordinates.  Our method hinges on the fact that the transformation
from the $(x,y)$ coordinates to the $(\widetilde X,\widetilde Y)$ does 
two things:

\noindent 1. it 
preserves area (this can be easily checked)

\noindent and 

\noindent 2. it straightens out the isotherms, i.e., transforms 
them to
the family of lines parallel to the $\widetilde Y$-axis in 
the $(\widetilde X,\widetilde Y)$-plane. 

Once we have this transformation we can reduce
the general case to the latter one which is much simpler.
For a more detailed and precise treatment we refer again to [Co].

The particular choice of the second coordinate  $Y=x$
probably requires some explanation.  What we need is a second family of curves
which is transversal to the isotherms.  We have chosen the simplest one---the
lines parallel to the $y$-axis.  This won't work in all situations but will for
most isotherms which arise in thermodynamics.  If there is a problem, one
can replace this family with the lines parallel to the $x$-axis
(the function $\Psi$ then has to be modified accordingly).  
At any given point, one of the two will work (at least locally).

We remark that it is a well-known fact of differential geometry that  
a transformation
with the above two properties 
is always possible---we can straighten out suitable families of curves 
by means of an area-preserving transformation. Our contribution has been to 
display a simple method
of constructing it explicitly and to point out its relevance for thermodynamics.

\section{The uncalibrated case}
Since the uncalibrated case is probably of less practical value, we discuss
it only briefly.  The situation here is that we are given the family
of isotherms but not their calibrations---they are specified as the
level curves of a function  $f(x,y)$ (e.g., empirical temperature).   
The function  $f$ is only determined up to a recalibration since the 
family of level curves cannot
distinguish
between $f$ and a function $F$ of the form $\phi \circ f$  where $\phi$
is a homeomorphism between intervals of the line.  In this case, if we 
have {\it two} adiabats we can once more recover the whole family of adiabats
and the {\it correct} calibration of  $f$.

In order to do this, we proceed as above to obtain the variables  $\widetilde X$ and $\widetilde Y$
which straighten out the isotherms.  The two adiabats are then transformed
into curves with equations  $\widetilde Y = f_0(\widetilde X)$ and 
$\widetilde Y = f_1(\widetilde X)$.  We  introduce the recalibration
$\phi(\widetilde X) =  \int (f_1(\widetilde X)-f_0(\widetilde X))
\,d\widetilde X$ and the 
level curves
of the function $$\frac{\widetilde Y-f_0(\widetilde X)}{f_1(\widetilde X)-f_0(\widetilde X)} $$
are the adiabats in the $(\widetilde X,\widetilde Y)$-plane.
Then the recalibrated temperature function is $\phi(f(x,y))$ and the
adiabats in the $(x,y)$-plane are the level curves of the functions
$$\frac{\widetilde Y(x,y)-f_0(\widetilde X(x,y))}{f_1(\widetilde X(x,y))-f_0(\widetilde X(x,y))} .$$ 

We illustrate this again with the example of the ideal gas.
We assume that the two adiabats are $x y^\gamma =1$ and $x y^\gamma = e$
and  we have mistakenly calibrated the isotherms not with $x y$ but
with $x^2 y^2$.
We will show how the knowledge of the two adiabats shows up the calibration
error and displays the correct version.

If we carry out the above scheme than we find that the transformation
has the form:

$$\widetilde X = x^2 y^2 ,\quad\widetilde Y = \frac 12 x y \ln x, $$
with inverse
$$x = \exp \left (\frac {2 \widetilde Y}{\widetilde X^{\frac 12}}\right ),\quad y=\widetilde X^{\frac12}\exp \left (\frac {-2 \widetilde Y}{\widetilde X^{\frac 12}}\right ) .$$
A simple computation shows that in this case:

$$f_0(\widetilde X)=\frac {\frac \gamma 2 \ln(\widetilde X)}{(\gamma-1)\widetilde X^{\frac 12}}, \quad
f_1(\widetilde X)=\frac {\frac \gamma 2 \ln(\widetilde X)-1}{(\gamma-1)\widetilde X^{\frac 12}} .$$
Hence their difference is $\dfrac 1{(\gamma-1)\widetilde X^{\frac 12}}$ with primitive the square root function
(up to a constant) and 
this determines the correct calibration of the temperature function.

\end{document}